\documentclass[aps,prl,toolkits,twocolumn]{revtex4}
\begin{document}

\title{QCD strings and the thermodynamics of the metastable phase of QCD
at large $N_c$}

\author{Thomas D. Cohen}
\email{cohen@physics.umd.edu}

\affiliation{Department of Physics, University of Maryland,
College Park, MD 20742-4111}


\begin{abstract}
The thermodyanmics of a metastable hadronic phase of QCD at large
$N_C$ are related to properties of an effective QCD string.  In
particular, it is shown that in the large $N_c$ limit and near the
maximum hadronic temperature, $T_H$, the energy density and pressure
of the metastable phase scale as ${\cal E} \sim
(T_H-T)^{-(D_\perp-6)/2}$ (for $D_\perp <6$) and $P \sim
(T_H-T)^{-(D_\perp-4)/2}$ (for $D_\perp <4$) where $D_\perp$ is the
effective number of transverse dimensions of the string theory. It
is shown, however,  that for the thermodynamic quantities of
interest the limits $T \rightarrow T_H$ and $N_c \rightarrow \infty$
do not commute.  The prospect of extracting $D_\perp$ via lattice
simulations of the metastable hadronic phase at moderately large
$N_c$ is discussed.
\end{abstract}

\maketitle

The modern incarnation of string theory as a fundamental
description of nature arose Phoenix-like out of the ashes of a
failed attempt to describe strong interaction dynamics in terms of
basic string degrees of freedom.  With the advent of QCD, it
rapidly became clear that a string model was not the correct way
to describe fully the theory of strong interactions.  However,
string models did have many successes in describing important
theoretical and phenomenological features of hadronic physics.  On
the phenomenological side, string models describe in a simple and
natural way the linearly rising Regge trajectories seen in the
data\cite{Pol}.  They also give rise to a spectrum of hadrons
which grow exponentially\cite{Pol} which is also seen in the
data\cite{spec,spec1}.  On the theory side such models naturally
have linear confinement. Intuitively this picture makes sense if
color sources form flux tubes---as they appear to on the lattice.
Accordingly from the early days of QCD on, there has been a
widespread belief that QCD in some sense becomes qualitatively
equivalent to an effective string theory, at least in a certain
domain.  Such an effective theory can be denoted as a``QCD
string''.

This paper relates the nature of the QCD string theory to its thermodynamics at large
$N_c$.  In the large $N_c$ limit QCD becomes a pure glue theory and has a first order
phase transition between its hadronic and deconfined phases\cite{Tep1}. The purpose
of this paper is to show that at large $N_c$ the thermodynamics of the metastable
hadron phase above this transition is determined by the nature of the string theory.
In the first place regardless of the type of string theory there is a maximum
temperature $T_H$ above which the metastable hadronic phase cannot be maintained.
Secondly, the power law behavior  of the energy density and the free energy of the
metastable hadronic phase is completely determined by the effective number of
transverse dimensions of the QCD string. This is significant because lattice studies
of the metastable phase of pure glue theory for relatively large $N_c$ (up to
$N_c=12$) are now possible\cite{Tep2}.

The qualitative successes of the string picture can only be rigorously correct
quantitatively in the large $N_c$ limit. For example, the linear Regge behavior and
the exponentially growing density of hadrons both depend on the identification of
hadron masses. However, the excited hadrons are resonances and thus do not have
well-defined masses.  Fortunately, as $N_c \rightarrow \infty$, the width of glueballs
and mesons goes to zero (as $N_c^{-2}$ and $N_c^{-1}$, respectively)\cite{largeN}, the
ambiguity in the value of the mass vanishes, and the string theory prediction becomes
well defined. Similarly, linear confinement only can occur in QCD if for some reason
flux tubes are unable to break.   In the large $N_c$ limit of QCD, a simple
diagrammatic argument shows that the formation of each quark-antiquark pair is
suppressed by a power of $1/N_c$ and thus at large $N_c$, flux tubes do not
break\cite{largeN}. Therefore, the simple consequences ascribed to a QCD string theory
can only accurately reflect the underlying theory in the large $N_c$ limit.

Similarly, one expects a string-like description to be increasingly more accurate as
the excitation energy increases.  If the QCD string theory is simulating a flux tube,
then one expects the dynamics of the tube moving collectively in the manner of a
string to arise when the characteristic length of the flux tube is much longer than
the typical width of the tube.  Such excitations have large masses.

For simplicity, the present analysis will be restricted to pure gauge theory.  This
restriction is motivated both by the focus on the large $N_c$ where quark loop
effects are suppressed by factors of $1/N_c$, and by the desire to compare the
analysis with lattice simulations which for relative large $N_c$ have only been done
for pure glue\cite{Tep2}.  The inclusion of quark loops leads to possible technical
complications in the analysis and will be neglected here in the interest of clarity.

Consider the behavior of large $N_c$ QCD at zero chemical potential in the hadronic
phase.  As argued in ref.~\cite {TDC} such a system is well described by a
non-interacting gas of hadrons (which for our pure glue system means glueballs).  The
hadrons are non-interacting since by standard $N_c$ counting agruments\cite{largeN}
glueball-glueball interactions scale with $N_c$ according to
 \begin{equation} V_{\rm g g} \sim   1/N_c^2 \; . \label{v}
 \end{equation}
  This scaling
implies that if the number densities of the hadrons are of order
$N_c^0$, then the energy is dominated by the mass of the hadrons
also of order $N_c^0$ and interactions may be neglected.

A framework for describing such a non-interacting gas of hadrons has been developed
for phenomenological reasons \cite{Kar} and will be used here. Consider a gas of
non-interacting glueballs where the $k^{\rm th}$ species of glueball has mass $m_k$
and spin $J_k$. It is straightforward to see that that the energy density
$\epsilon_k$ and pressure associated with each spin state of the boson is given by
\begin{eqnarray}
\epsilon(m_k,T)  &=& \int \frac{{\rm d}^3 p}{(2 \pi)^3} \, \frac{
\sqrt{p^2 + m_k^2}}{e^{\sqrt{p^2 + m_k^2}/T}-1} \nonumber \\
 & \rightarrow &  \frac{T^{3/2} \, m_k^{5/2} \,e^ {-
m_k/T}}{ (2 \pi)^{3/2}}\,=\, m_k \, n_k \, , \nonumber \\
P(m_k,T)& = &\left ( \frac{m_k^2 T^2}{2 \pi^2} \right ) \,
\sum_{n=1}^\infty \, K_2 \left(\frac{n m_k}{T} \right) \nonumber \\
&\rightarrow& \frac{ m^{3/2} T^{5/2}}{2^{3/2} \pi^{3/2}}\,
e^{-m/T} \; \label{f}
\end{eqnarray}
where $K_2$ is a modified Bessel function and the arrows indicate
asymptotic behavior of these functions as $ m/T \rightarrow \infty
$.  The spectral density of hadrons is defined as $ \rho(m) \equiv
\sum_k (2 J_k+1 )\, \delta(m - m_k) $.  The asymptotic
contribution to energy density and the pressure in the hadronic
phase are given by
\begin{eqnarray}
 {\cal E}(T)&=& \int_0^\infty {\rm d} m \, \rho (m)
\epsilon(m,T) + {\cal O}(1/N_c^2) \nonumber \\ \,  \label{tot}
\end{eqnarray}
where the $ {\cal O}(1/N_c^2)$ correction arises from the interaction energy between
the glueballs.  The specific heat can be obtained by differentiating the energy
density with respect to the temperature.  Generically both the energy density and the
hadron number density are of order $N_c^0$.  At low temperatures and large $N_c$, the
energy density is dominated by the mass of the hadrons rather than their interaction
energy; QCD can then be described increasingly accurately in terms of non-interacting
hadrons as $N_c$ increases.

However, at very high temperatures QCD is described increasingly
accurately as a gas of weakly interacting quarks and gluons.  At
large $N_c$ this is dominated by the gluons and the energy density
scales as $N_c^2$. Moreover, from the behavior of the $\beta$
function it is apparent that the characteristic temperature at
which the thermodynamics is well described in terms of weakly
interacting quarks and gluons is of order $N_c^0$. Thus, there are
two temperature domains at order $N_c^0$: a low temperature
hadronic one with energy densities of order $N_c^0$ and a
quark-gluon phase with energy densities of order $N_c^2$. For
infinite $N_c$ such a situation is only possible if there is a
discontinuity: below some $T_c$, ${\cal E}/N_c^2 =0$ while above
$T_c$ ${\cal E} /N_c^2$ is nonzero. In fact, for pure glue
theories we know from lattice studies that this transition is
first order. Moreover, the latent heat of the transition appears
to scale numerically with $N_c^2$ as expected from this simple
analysis\cite{Tep2}.

Suppose that the high-lying spectrum of hadrons in large $N_c$ QCD is, in fact,
describable as a hadronic string.  The behavior of the density of hadronic states is
known for asymptotically large values of $M$ \cite{Dien} and is given by
 \begin{equation}
 \rho(m) \rightarrow \rho_{\rm asy}(m)= A \, (m/T_H)^{-2 B} \, e^{m/T_H}
 \label{stringrho} \end{equation}
where $A$, $T_H$ and $B$ are parameters which are independent of $N_c$.  The value of
the $T_H$ depends on the numerical value of the string tension $\sigma$. It also
depends on the type of string theory and, in particular, the world sheet central
charge, c: $T_H^2 = \frac{3 \sigma}{\pi c}$.  The key to the present analysis is the
parameter $B$  is universal\cite{Dien} ({\it i.e.}, independent of the type of string
theory) and depends only on, $D_\perp$;
 \begin{equation}
 B  =  (D_\perp+3)/4  \; .
 \end{equation}

The parameter $T_H$ is the well-known Hagedorn temperature associated with the
spectrum\cite{Hag}. In the  hadronic gas model, the energy density in Eq.~(\ref{tot})
diverges for any $T>T_H$. With the advent of QCD it became apparent that the Hagedorn
temperature was the upper bound for {\it hadron} matter\cite{Cab}; the system is
necessarily in a deconfined phase above $T_H$. As we now know, large $N_c$ QCD enters
a deconfined phase {\it before} $T_H$ is reached\cite{Tep1}. However, $T_H$
presumably represents an upper bound for the temperature of a {\it metastable}
hadronic phase\cite{Tep2}.

It was realized long ago that the power law behavior of thermodynamic quantities in
the neighborhood of $T_H$ are determined by the power law of the prefactor in the
Hagedorn expression; that is, by the coefficient $B$\cite{HW}. Evaluating the
integrals for the asymptotic contributions to the energy density and pressure using
the Hagedorn spectrum obtained from QCD strings yields:
\begin{widetext}
\begin{eqnarray}
{\cal E}_{\rm asy}  \, & =  \,(2 \pi)^{-3/2} \,  A \,
T^{(9-D_\perp)/2} \, \, T_H^{7/2}\, \times    \left \{
\begin{array}{cc}
 (T_H-T)^{(D_\perp-6)/2} \, \Gamma \left (\frac{6-D_\perp}{2} \right
)  &\text{ if  $D_\perp <6$} \\ & \\
 -\log \left ((T_H - T)/T_H) \right )& \text{if $D_\perp = 6$}
\end{array}
 \right. \nonumber \\ \nonumber\\ \nonumber\\
{P}_{\rm asy}  \, & = \, (2 \pi)^{-3/2} \, A \, T^{(9-D_\perp)/2}
\, \, T_H^{5/2}\, \times    \left \{ \begin{array}{cc}
 (T_H-T)^{(D_\perp-4)/2} \, \Gamma \left (\frac{4-D_\perp}{2} \right
)  &\text{ if  $D_\perp <4$}
\\ & \\  -\log \left ((T_H - T)/T_H) \right )& \text{ if $D_\perp = 4$}
\end{array}
 \right . \; .  \label{pl}
\end{eqnarray}
\end{widetext}
Since the asymptotic contributions to ${\cal E}$ and $P$ diverge in the neighborhood
of $T_H$, they will dominate the full expressions which in turn will diverge in the
same manner. The nonanalytic behavior of the energy density in the vicinity of $T_H$
immediately implies nonanalyticity for the specific heat. Quite generally one sees
that
\begin{equation}
\alpha = 4 - D_\perp/2 \label{alpha} \; ,
\end{equation}
where $\alpha$ is analogous to the usual critical exponent for the
specific heat ($c_v \sim (T_H-T)^\alpha$).

Suppose, that the QCD string has $D_\perp \le 6$. The thermodynamic behavior of such
a system is qualitatively quite different from what one sees in typical statistical
mechanics models---at least at large $N_c$. For typical models the metastable phase
can have a maximum temperature. Such a maximum temperature is a spinodal point---the
point at which local fluctuations destabilize the phase.  The energy density remains
finite as the spinodal point is approached---although the specific heat will
diverge.  In this sense, a spinodal point behaves essentially as a second phase order
transition.  If the QCD string has $D_\perp \le 6$, there is also a maximum
temperature for the metastable phase, namely, $T_H$, as there is with familiar
statistical mechanical models. However, in sharp contrast to the typical spinodal
point, the energy density of the QCD string theory {\it diverges} as $T_H$ is
approached; for the case of $D_\perp=2$ one sees from Eq.~(\ref{pl}) that the power
law behavior for the energy density of the metastable hadronic phase goes as ${\cal
E} \sim (T_H-T)^{-2}$.  This inability to reach the maximum temperature in the
hadronic phase due to its energetic inaccessibility is the modern incarnation of
Hagedorn's old idea (although now relegated to a metastable phase and large $N_c$)
and occurs for $D_\perp \le 6$. The critical exponent $\alpha$ encodes the difference
between typical statistical systems for which the finiteness of the energy at the
transition requires $0 \le \alpha <1$ and the present case for which $\alpha$ is
greater than unity (provided $D_\perp < 6$). For example in the simple Nambu-Goto
string ($D_\perp=2$) $\alpha=3$ .

Lattice simulations showing clear evidence of the energy density
having such a power law divergent behavior for the energy density
at large $N_c$ would provide strong numerical support for the
notion of the QCD string. Similarly a numerical determination that
$\alpha =3$ for the metastable phase  would go a long way toward
establishing that the effective QCD string has $D_\perp=2$ as in
the Nambu-Goto string.

Perhaps more exciting is the possibility that numerical evidence could show that
$\alpha \ne 3$ near $T_H$ indicating the effective transverse dimension of the QCD
string is different from two.  The effective transverse dimension of the QCD string
could, in principle, be different from the number of transverse space-time dimensions
since the effective number of transverse dimensions characterizes the number of
degrees of freedom of the string.  One can imagine scenarios in which the dynamics
gives  a larger number of degrees of freedom than in the simplest string picture.  As
a practical matter, properties of the QCD string can be extracted from the behavior
of the system near the endpoint of the metastable phase only if $D_\perp \le 6$. For
$D_\perp > 6$, the  energy density does not diverge at $T_H$ and it becomes difficult
to disentangle the Hagedorn behavior of string theory from the critical behavior seen
in many statistical mechanical systems.

An important caveat is in order.  The preceding analysis makes important use of the
large $N_c$ limit.  One might hope that the qualitative behavior of power law
divergence encoded in Eq~(\ref{pl}) remains valid at large but finite $N_c$ (perhaps
with the powers altered by $1/N_c$ corrections).  Unfortunately, for $D_\perp \le 4$
simple thermodynamics indicates that this cannot be the case. The pressure is the
negative of the free energy density. By definition in a metastable phase there exists
another phase with lower free energy (and thus higher pressure). In this case it is
the deconfined phase (with pressure $\sim N_c^2$). Thus, the expression given for the
pressure in Eq.~(\ref{pl}) breaks down when $T$ gets sufficiently close to $T_H$ so
that pressures are of order $N_c^2$. This breakdown can be understood
physically---the analysis is based on negligible interactions between glueballs.
However, as $T \rightarrow T_H$ the glueball density diverges. It is easy to show
that the number density of glueballs scales in the same way as the pressure (at large
$N_c$) and thus one expects that when the pressures reach order $N_c^2$ so do the
number densities; Eq.~(\ref{v}) then implies that the interaction energy does as
well. Thus at that point the interaction energy is comparable  to the mass energy and
analysis breaks down.

It is clear that  the temperature regime in which the expressions in Eq.~(\ref{pl})
dominate, the energy and the pressure is limited. For low temperatures, the
asymptotic form does not dominate, while very close to $T_c$ hadron-hadron
interactions invalidate the analysis (provided $D_\perp \le 4$). Evidently the domain
of validity is
\begin{equation}
c_1 T_H \ll T_H-T \ll {T_H} \, \times \left \{ \begin{array}{cc}
  c_2 {N_c^{-\frac{4}{4-D_\perp}}} &\text{ if   $D_\perp <4$}
\\ & \\ e^{-c_2 N_c^2} & \text{if $D_\perp = 4$} \end{array} \right
.
\end{equation}
where $c_1$ and $c_2$ are constants of order $N_c^0$. Thus for the
simple scalar string with $D_\perp=2$, the range of validity $c_1
T_H \ll T_H-T \ll {T_H}/N_c^2$.

Note that as $N_c \rightarrow \infty$ the upper end of the region
of validity approaches $T_H$.  However, for any finite value of
$N_c$ the expression ultimately must break down as $T$ approaches
$T_H$. Thus the critical exponents for the infinite $N_c$ theory
will differ substantially from the critical exponents for large
but finite $N_c$: the limit of the critical exponent as $N_C
\rightarrow \infty$ will not equal the critical exponent for
infinite $N_c$. Recall the critical exponent associated with some
quantity $f(T)$ that diverge at the critical point is defined as
a limit:  $\gamma_f \equiv \lim_{T \rightarrow T_c} \left
((T_c-T) \frac{d \, \log \left ( f(T) \right ) }{d \, T} \right
)$. However, for the quantities of interest here the limits $N_c
\rightarrow \infty$ and $T \rightarrow T_H$ are not uniform.

The nonuniformity of limits means that in extracting the power law behavior to learn
about hadronic strings, one considers the $N_c \rightarrow \infty$ limit before the $T
\rightarrow T_H$ limit.  Care must be taken in extracting the power law behavior of
the energy density that one is working in the regime of validity of Eq.~(\ref{pl}),
{\it i.e.}, for temperatures not too close to $T_H$.  Since in practice it is very
difficult numerically to study a metastable phase very near its endpoint, it may not
be too difficult to do this provided $N_c$ is large enough.

The study of $N_c$ scaling behavior makes manifest the fact the range of validity of
Eq.~(\ref{pl}) is limited at the top end. This helps resolve a very old problem with
Hagedorn's original idea of a limiting temperature.   The issue of whether
hadron-hadron interactions could fundamentally alter the thermodynamics was
recognized as important very early by Huang and Weinberg\cite{HW}.  However, at the
time there were no techniques to study the issue in a controlled manner.  In practice
it only could be handled in an {\it ad hoc} fashion by neglecting the interactions
under the assumption that such interactions did not fundamentally alter the physical
density of states. The $N_c$ scaling arguments for the pressure show that this is not
valid for the real world of $N_c=3$ or, indeed, for any finite $N_c$---there will
necessarily be a regime in which the hadron-hadron interactions must cause the growth
in the pressure to saturate as $T$ approaches  $T_H$ while the thermodynamics of a
non-interacting gas of hadrons with a Hagedorn spectrum will have a diverging
pressure.

The preceding analysis provides no real insight into the behavior of the system at
finite (but large) $N_c$  in the immediate vicinity of $T_H$. A highly plausible
scenario is that there still is a maximum hadronic temperature at a spinodal point
$T_c$ which is very close to $T_H$: ($(T_H -T_c)/T_H) \sim N_c^2$.  As $T_c$ is
approached glueball-glueball interactions become strong and this drives an
instability leading to the existence of a spinodal point.  As was argued earlier, for
the case of $D_\perp \le 4$ the breakdown in the hadron gas analysis occurs at
temperatures where the energy of glueball-glueball interactions become comparable to
the energies of the free particles; at this point, the dynamics changes
qualitatively. Thus, when pressure of order $N_c^2$ are reached, nothing protects the
system from developing an instability.

The relationship of the nature of the QCD string  thermodynamics
of the metastable phase of large $N_c$ pure Yang-Mills theory
would be of comparatively little importance if there were no
independent way to study this thermodynamics.  However, lattice
simulations of the metastable phase provide the possibility of
direct calculations of the thermodynamics from QCD.  Simulations
of the metastable phase at rather large $N_c$ have already been
done\cite{Tep2}. These studies were aimed at identifying the
maximum hadronic temperature (identified as the Hagedorn
temperature) which was done by extrapolating calculated values of
the string tension down to the point at which it vanishes.  If
similar calculations were done for thermodynamic quantities such
as the pressure, the energy density or the specific heat,  it may
be possible to get {\it ab initio} evidence of the nature of the
QCD string.

{\it Note added after publication.}  After this paper was
published, the author became aware of an important early paper by
Charles Thorn on the subject of the QCD phase transition at large
$N_c$ and its relation to the Hagedorn spectrum\cite{CT}.
Reference \cite{CT} concludes among other things that QCD must
have a phase transition at infinite $N_c$ and that the Hagedorn
temperature is bounded from below by the critical temperature for
the transition. Given the pioneering nature of this paper, and the
fact that it is not as widely known as it should be, it is
important to cite it here.

\acknowledgments{The author is grateful to S. Nussinov for useful comments.  Support
of the US DOE under grant number DE-FG02-93ER-40762 is acknowledged.}

\end{document}